\documentclass[preprint,10pt,pra,notitlepage,twocolumn,superscriptaddress]{revtex4}%
\usepackage{amsfonts}
\usepackage{amsmath}
\usepackage{amssymb}
\usepackage{graphicx}%
\providecommand{\U}[1]{\protect\rule{.1in}{.1in}}
\providecommand{\U}[1]{\protect\rule{.1in}{.1in}}

\begin{document}
\preprint{ }
\title{Superfluid pairing in a mixture of a spin-polarized Fermi gas and a dipolar condensate}
%\author{Ben Kain$^{1}$ and Hong Y. Ling$^{2}$}
%\affiliation{$^{1}$Department of Physics, College of the Holy Cross, Worcester, MA 01610}
%\affiliation{$^{2}$Department of Physics and Astronomy, Rowan University, Glassboro, New Jersey 08028}
\author{Ben Kain}
\affiliation{Department of Physics, College of the Holy Cross, Worcester, MA 01610, USA}
\author{Hong Y. Ling}
\affiliation{Department of Physics and Astronomy, Rowan University, Glassboro, NJ 08028, USA}

\begin{abstract}
\noindent We consider a mixture of a spin-polarized Fermi gas and a dipolar
Bose-Einstein condensate in which s-wave scattering between fermions and the
quasiparticles of the dipolar condensate can result in an effective attractive
Fermi-Fermi interaction anisotropic in nature and tunable by the dipolar
interaction. \ We show that such an interaction can significantly increase the
prospect of realizing a superfluid with a gap parameter characterized with a
coherent superposition of all odd partial waves. We formulate, in the spirit
of the Hartree-Fock-Bogoliubov mean-field approach, a theory which allows us
to estimate the critical temperature when the anisotropic Fock potential is
taken into consideration and to determine the system parameters that optimize
the critical temperature at which such a superfluid emerges before the system
begins to phase separate.

\end{abstract}
\maketitle

\section{Introduction}

Superfluid pairing of fermions in the $l$th partial wave depends on the
two-body scattering amplitude $f_{l}$. \ According to Wigner's threshold laws
\cite{landau89}, $f_{l}$ scales as $\left(  k_{F}^{0}\right)  ^{2l}$ for
typical ground state atoms, where $\hbar k_{F}^{0}$ is the Fermi momentum.
This reduces the critical temperature for the p-wave ($l=1$)
Bardee-Cooper-Schrieffer (BCS) superfluid to such a low level that it is
virtually inaccessible to current technology, except in situations where
Wigner's threshold laws are not respected. \ A case in point is tuning atoms
near the p-wave Feshbach resonance \cite{bohn00,klinkhamer04,ohashi05,ho05},
where p-wave scattering can be dramatically enhanced. \ This excited the hope
of realizing $p$-wave superfluids via p-wave Feshbach resonances in the
ultracold atomic physics community.\ But, due to an increased number of
inelastic collisions near the resonance, \textquotedblleft Feshbach
molecules\textquotedblright\ are short lived
\cite{regal03,zhang04,ticknor04,gunter05}, and such a prospect appears
difficult to attain. \ Another example, first pointed out by You and Marinescu
\cite{marinescu98}, is subjecting a dipolar Fermi gas to a sufficiently strong
DC electric field, where p-wave scattering can also be enhanced. \ A
subsequent detailed investigation by Baranov et al. \cite{baranov02} indicates
that the order parameter in such a spin-polarized dipolar Fermi gas is unusual
in the sense that it is the superposition of all odd partial waves. \ Recently
there has been an upsurge of effort aimed at achieving unusual Fermi pairings
in both two-component\ and single component dipolar Fermi gases
\cite{samokhin06,bruun08, wu10, shi10,liao10,ronen10,zhao10} along the
direction of You and Marinescu \cite{marinescu98}, \ motivated largely by
recent rapid experimental advancement in achieving ultracold dense dipolar
gases in $^{40}$K -$^{87}$Rb polar molecules
\cite{ospelkaus08,ni08,ospelkaus09} and in Cr \cite{stuhler05}, spin-1 Rb
\cite{vengalattore08}, and Dy \cite{lu11} atoms.

In the present work, we focus on an alternative route towards the same goal,
seeking to engineer unusual superfluid pairings in mixtures of different
degenerate quantum gases. Cold-atom systems, besides being capable of
unprecedented controllability (interaction strength, dimensionality, etc.),
enjoy the distinct advantage over traditional solid-state systems of being
easily mixed to form new states of quantum gases and liquids. \ Bose-Bose,
Bose-Fermi, and Fermi-Fermi mixtures of different isotopes or species are now
readily accessible under (sophisticated) laboratory conditions (see
\cite{shin08,park11,wu11} and references therein). As is well known, due to
the density fluctuation of Bose condensates, mixing bosons can induce an
effective attraction between two fermions \cite{bijlsma00}. \ This was the
mechanism in Efremov and Viverit's proposal \cite{efremov02} to achieve p-wave
Cooper pairings in a single-component Fermi-Bose mixture. \ A generalization
of the scheme to a two-dimensional (2D) mixture involving dipolar bosons was
recently carried out by Dutta and Lewenstein \cite{dutta10}, with the goal of
realizing a superfluid with $p_{x}+ip_{y}$ symmetry whose excitations are
non-Abelian anyons that are the building blocks for topological quantum
computation \cite{nayak08}. \ Nishida \cite{nishida09} and Nishida and Tan
\cite{nishida10} sought to achieve the same goal by inducing p-wave and higher
partial-wave resonances from s-wave Feshbach resonance between atoms from
different Fermi gases in different dimensions. In a recent article
\cite{kain11}, we explored singlet and triplet superfluid competition in a
mixture of two-component Fermi and one-component dipolar Bose gases.

In this paper, we consider a three-dimensional (3D) mixture between
(nondipolar) fermions of mass $m_{F}$ in ground state $\left\vert
a\right\rangle $ and (dipolar) bosons of mass $m_{B}$ in ground state
$\left\vert b\right\rangle $, with their dipoles oriented along the $z$
direction established by an external (either electric or magnetic) field. \ In
contrast to the phonon spectrum of a nondiploar condensate, which is
isotropic, the phonon spectrum of a dipolar condensate is anisotropic
\cite{goral00,giovanazzi02}. \ Such anisotropic phonons will thus induce an
anisotropic Fermi-Fermi interaction, thereby providing an alternative
mechanism for creating a superfluid (in the mixture between a spin polarized
Fermi gas and a dipolar condensate) with an order parameter which is the
superposition of all odd partial waves.

The outline of our paper is as follows. \ In Sec.
\ref{Sec:theoretical Formulation}, we develop, within the framework of the
Hartree-Fock-Bogoliubov mean-field approach, a theoretical foundation for
self-consistently addressing challenges we will face in later sections.
\ These challenges include the anisotropy of the induced interaction and the
limitations due to phase separation. \ In Sec. \ref{Sec:surface deformation},
we calculate the critical temperature taking into consideration the
anisotropic Fock potential with the help of the variational method, and we
show that properly tuning the dipolar interaction can dramatically increase
the critical temperature in the long wavelength limit. \ \ In Sec.
\ref{Sec:optimization}, we review the subject of phase separation and develop
a procedure that allows us to determine the system parameters for achieving
the optimal critical temperature before the system begins to phase separate.
\ We summarize and conclude in Sec. \ref{Sec:conclusion}.

\section{Theoretical Formulation}

\label{Sec:theoretical Formulation}

\subsection{Bare Model}

In our model the low temperature physics arises from the interplay between
short- and long-range two-body interactions. \ The short-range interaction is
dominated by s-wave collisions which, given that the Pauli exclusion principle
forbids such interactions taking place between two identical fermions, are
parameterized with two scattering lengths,\ $a_{BB}$ and $a_{BF}$,
characterizing s-wave scattering between two bosons and between a boson and a
fermion, respectively. \ The long-range interaction is the dipole-dipole
interaction (restricted only to bosons in our model), which varies with
$\mathbf{r}$, the separation between two particles in position space,
according to $U_{DD}\left(  \mathbf{r}\right)  =d^{2}\left(  1-3z^{2}%
/r^{2}\right)  /r^{3}$, where $d^{2}$ represents the dipolar interaction
strength. \ \ Assuming that the mixture is sufficiently large so that it can
be treated as a uniform system with an effective volume $V$, we model such a
mixture with a (grand canonical) Hamiltonian $\hat{H}=\hat{H}_{B}+\hat{H}%
_{F}+\hat{H}_{BF}$ in momentum space according to
\begin{align}
\hat{H}_{B}  &  =\sum_{\mathbf{k}}\left(  \epsilon_{\mathbf{k},B}%
-u_{B}\right)  \hat{b}_{\mathbf{k}}^{\dag}\hat{b}_{\mathbf{k}}+\left(
2V\right)  ^{-1}\times\nonumber\\
&  \sum_{\mathbf{k},\mathbf{k}^{\prime},\mathbf{q}}\left[  U_{BB}%
+U_{DD}\left(  \mathbf{q}\right)  \right]  \hat{b}_{\mathbf{k}+\mathbf{q}%
}^{\dag}\hat{b}_{\mathbf{k}^{\prime}-\mathbf{q}}^{\dag}\hat{b}_{\mathbf{k}%
^{\prime}}\hat{b}_{\mathbf{k}},\label{HB}\\
\hat{H}_{F}  &  =\sum_{\mathbf{k},\sigma}\left(  \epsilon_{\mathbf{k},F}%
-u_{F}\right)  \hat{a}_{\mathbf{k}}^{\dag}\hat{a}_{\mathbf{k}},\\
\hat{H}_{BF}  &  =U_{BF}V^{-1}\sum_{\mathbf{k},\mathbf{k}^{\prime},\mathbf{q}%
}\hat{a}_{\mathbf{k}}^{\dag}\hat{a}_{\mathbf{k}+\mathbf{q}}\hat{b}%
_{\mathbf{k}^{\prime}}^{\dag}\hat{b}_{\mathbf{k}^{\prime}-\mathbf{q}},
\label{HBF}%
\end{align}
where $\hat{b}_{\mathbf{k}}$ ($\hat{a}_{\mathbf{k}}$) is the field operator
for annihilating a boson (a fermion) with kinetic energy $\epsilon
_{\mathbf{k},B}=\hbar^{2}k^{2}/2m_{B}$ $\left(  \epsilon_{\mathbf{k},F}%
=\hbar^{2}k^{2}/2m_{F}\right)  $ and chemical potential $u_{B}$ ($u_{F}$);
$U_{BB}=4\pi\hbar^{2}a_{BB}/m_{B}$ and $U_{BF}=4\pi\hbar^{2}a_{BF}/m_{BF}$
with $m_{BF}=2m_{B}m_{F}/(m_{B}+m_{F})$ measure the related s-wave scattering
strengths; $U_{DD}\left(  \mathbf{k}\right)  =8\pi d^{2}P_{2}\left(
\cos\theta_{\mathbf{k}}\right)  /3$ is the dipole-dipole interaction in
momentum space, with $P_{2}\left(  x\right)  =\left(  3x^{2}-1\right)  /2$ the
second-order Legendre polynomial and $\theta_{\mathbf{k}}$ ($\phi_{\mathbf{k}%
}$) the polar (azimuthal) angle of vector $\mathbf{k}$.

\subsection{Effective (or Dressed) Model}

We consider the low temperature regime where virtually all bosons are
condensed to the zero-momentum mode so that we can apply the\ usual Bogoliubov
ansatz, in which $\hat{b}_{\mathbf{k}=0}$ is treated as a c-number
$b_{\mathbf{k}=0}$. \ This transforms Eq. (\ref{HB}) into a diagonalized form
with the familiar interpretation that the bosonic system described by Eq.
(\ref{HB}) is equivalent to a homogeneous dipolar condensate with density
$n_{B}=\left\vert b_{\mathbf{k}=0}\right\vert ^{2}$ plus a collection of
phonon modes that obey the\ Bogoliubov dispersion relation
\cite{goral00,giovanazzi02}%
\begin{equation}
E_{\mathbf{k}}=v_{B}\hbar k\sqrt{1+\left(  \xi_{B}k\right)  ^{2}%
+2\varepsilon_{dd}P_{2}\left(  \cos\theta_{\mathbf{k}}\right)  },
\end{equation}
where $\varepsilon_{dd}=4\pi d^{2}/(3U_{BB})$ \cite{dell04} measures the
dipolar interaction relative to the s-wave collision, $v_{B}=\sqrt{n_{B}%
U_{BB}/m_{B}}$ is the phonon speed, and $\xi_{B}$ $=\hbar/\sqrt{4m_{B}%
n_{B}U_{BB}}$ is the healing length. \ As can be seen, when $\varepsilon_{dd}$
$>$ $1$, phonons with $k\rightarrow0$ acquire imaginary frequencies. In our
study, then, we will limit $\ \varepsilon_{dd}$ \ to be less than one so that
the dipolar condensate is stable against collapse. \ Integrating out the
phonon degrees of freedom as in Ref. \cite{bijlsma00} leads to an effective
Bose-Fermi model described by the Hamiltonian%
\begin{equation}
\hat{H}^{E}=H_{B}^{0}+\hat{H}_{F}^{E}, \label{H Effective}%
\end{equation}
where%
\begin{equation}
H_{B}^{0}=V\left(  \frac{1}{2}U_{BB}n_{B}^{2}-u_{B}n_{B}\right)  \label{HB0}%
\end{equation}
accounts for the condensate energy, and
\begin{align}
\hat{H}_{F}^{E}  &  =\sum_{\mathbf{k}}\left(  \epsilon_{\mathbf{k},F}%
-u_{F}+U_{BF}n_{B}\right)  \hat{a}_{\mathbf{k}}^{\dag}\hat{a}_{\mathbf{k}%
}+\nonumber\\
&  \frac{1}{2V}\sum_{\mathbf{k},\mathbf{k}^{\prime},\mathbf{q}}U\left(
\mathbf{q}\right)  \hat{a}_{\mathbf{k}+\mathbf{q}}^{\dag}\hat{a}%
_{\mathbf{k}^{\prime}-\mathbf{q}}^{\dag}\hat{a}_{\mathbf{k}^{\prime}}\hat
{a}_{\mathbf{k}}, \label{HEF}%
\end{align}
is the effective Hamiltonian for the dressed fermions. In Eq. (\ref{HEF}),
\begin{equation}
U\left(  \mathbf{k}\right)  =\frac{U\left(  0\right)  }{1+\left(  \xi
_{B}k\right)  ^{2}+2\varepsilon_{dd}P_{2}\left(  \cos\theta_{\mathbf{k}%
}\right)  } \label{Uind}%
\end{equation}
is the phonon-induced Fermi-Fermi interaction in the static limit
\cite{bijlsma00,efremov02}.\ It is well known that the dipolar interaction in
position space is zero when averaged over all directions or, equivalently, the
dipolar interaction in momentum space, $U_{DD}\left(  \mathbf{k}\right)  $,
satisfies $U_{DD}\left(  \mathbf{k}=0\right)  =0$ and thus $U_{DD}\left(
\mathbf{k}\right)  $ is non analytic when $\mathbf{k}$ approaches zero
\cite{chan10}. \ For the same reason, the induced interaction $U\left(
\mathbf{k}\right)  $ is also non analytic when $\mathbf{k}$ approaches zero,
but instead of being zero, $U\left(  \mathbf{k}=0\right)  $ approaches the
finite value%
\begin{equation}
U\left(  0\right)  =-U_{BF}^{2}/U_{BB}\text{,}%
\end{equation}
which is certainly different from $U\left(  \mathbf{k}\rightarrow0\right)  $
according to Eq. (\ref{Uind}) due to its dependence on the polar angle in the
denominator. \ As can be seen, the induced interaction depends on the dipole
orientation differently than the direct dipole-dipole interaction and
therefore provides an alternative mechanism for the exploration of the physics
behind the anisotropy-related unusual Fermi pairings.

\subsection{Hartree-Fock-Bogoliubov Mean-Field Theoretical Treatment}

To study the superfluidity of a fermionic system described by the effective
Hamiltonian (\ref{HEF}), we use the self-consistent Hartree-Fock mean-field
approach \cite{gennes89} in which we pair fermionic field operators in the
two-body interaction part of the Hamiltonian in Eq. (\ref{HEF}) according to
Wick's theorem, associating direct pairing with the Hartree potential%
\begin{equation}
\Sigma^{^{\prime}}=U\left(  0\right)  n_{F}\text{,} \label{hartree}%
\end{equation}
exchange pairing with the Fock potential
\begin{equation}
\Sigma\left(  \mathbf{k}\right)  =-\frac{1}{V}\sum_{\mathbf{k}^{\prime}%
}U\left(  \mathbf{k}-\mathbf{k}^{\prime}\right)  \left\langle \hat
{a}_{\mathbf{k}^{\prime}}^{\dag}\hat{a}_{\mathbf{k}^{\prime}}\right\rangle ,
\label{Fock}%
\end{equation}
and BCS pairing with the gap parameter between fermionic pairs of opposite
momenta \
\begin{equation}
\Delta\left(  \mathbf{k}\right)  =\frac{1}{V}\sum_{\mathbf{k}^{\prime}}%
U_{A}\left(  \mathbf{k},\mathbf{k}^{\prime}\right)  \left\langle \hat
{a}_{-\mathbf{k}^{\prime}}\hat{a}_{\mathbf{k}^{\prime}}\right\rangle ,
\label{Gap}%
\end{equation}
where $n_{F}$ is the Fermi number density. \ In arriving at Eq. (\ref{Gap}),
we used a symmetrized induced interaction%
\begin{equation}
U_{A}\left(  \mathbf{k},\mathbf{k}^{\prime}\right)  =\frac{1}{2}\left[
U\left(  \mathbf{k}-\mathbf{k}^{\prime}\right)  -U\left(  \mathbf{k}%
+\mathbf{k}^{\prime}\right)  \right]  , \label{U(k,k')}%
\end{equation}
in order to explicitly restrict the gap parameter to the sector in which
$\Delta\left(  \mathbf{k}\right)  =-\Delta\left(  -\mathbf{k}\right)  $, a
constraint imposed by Fermi statistics (or equivalently the Pauli exclusion
principle). \ This means that $\Delta\left(  \mathbf{k}\right)  $ will be a
superposition involving all odd partial waves when the Fermi-Fermi interaction
is anisotropic \cite{kain11}, as is the case with our model here .

The effective Hamiltonian in Eq. (\ref{H Effective}) now becomes the
mean-field Hamiltonian
\begin{equation}
\hat{H}^{M}=H_{B}^{0}+H_{F}^{0}+\hat{H}_{F}^{^{M}}, \label{mean-field HF}%
\end{equation}
where $H_{B}^{0}$ has already been defined in Eq. (\ref{HB0}), and the other
two are defined, in terms of the self-consistent mean fields introduced in
Eqs. (\ref{hartree}) - (\ref{Gap}), as
\begin{align}
\hat{H}_{F}^{M}  &  =\sum_{\mathbf{k}}\left[  \xi_{\mathbf{k}}=\epsilon
_{\mathbf{k},F}-\mu_{F}+\Sigma\left(  \mathbf{k}\right)  \right]  \hat
{a}_{\mathbf{k}}^{\dag}\hat{a}_{\mathbf{k}}\nonumber\\
+  &  \frac{1}{2}\sum_{\mathbf{k}}\left[  \Delta\left(  \mathbf{k}\right)
\hat{a}_{\mathbf{k}}^{\dag}\hat{a}_{-\mathbf{k}}^{\dag}+h.c\right]  ,
\label{HFM}%
\end{align}
where $\mu_{F}$ is an effective chemical potential for fermions defined by%
\begin{equation}
\mu_{F}=u_{F}-\left[  U\left(  0\right)  n_{F}+U_{BF}n_{B}\right]  ,
\label{mu}%
\end{equation}
and
\begin{align}
&  H_{F}^{0}=-V\frac{1}{2}U\left(  0\right)  n_{F}^{2}\nonumber\\
&  -\frac{1}{2}\sum_{\mathbf{k}}\Sigma\left(  \mathbf{k}\right)  \left\langle
\hat{a}_{\mathbf{k}}^{\dag}\hat{a}_{\mathbf{k}}\right\rangle -\frac{1}{2}%
\sum_{\mathbf{k}}\Delta\left(  \mathbf{k}\right)  \left\langle \hat
{a}_{\mathbf{k}}^{\dag}\hat{a}_{-\mathbf{k}}^{\dag}\right\rangle , \label{HF0}%
\end{align}
or equivalently
\begin{align}
H_{F}^{0}  &  =-V\frac{1}{2}U\left(  0\right)  n_{F}^{2}\nonumber\\
&  +\frac{1}{2}\sum_{\mathbf{k},\mathbf{k}^{\prime}}\Sigma\left(
\mathbf{k}\right)  U^{-1}\left(  \mathbf{k}-\mathbf{k}^{\prime}\right)
\Sigma\left(  \mathbf{k}^{\prime}\right) \nonumber\\
&  -\frac{1}{2}\sum_{\mathbf{k},\mathbf{k}^{\prime}}\Delta\left(
\mathbf{k}\right)  U_{A}^{-1}\left(  \mathbf{k},\mathbf{k}^{\prime}\right)
\Delta^{\ast}\left(  \mathbf{k}^{\prime}\right)  , \label{HF0 Inverse}%
\end{align}
where the inverse matrices are defined as $V^{-1}\sum_{\mathbf{k}%
^{\prime\prime}}U\left(  \mathbf{k}-\mathbf{k}^{\prime\prime}\right)
U^{-1}\left(  \mathbf{k}^{\prime\prime}-\mathbf{k}^{\prime}\right)
=\delta_{\mathbf{k},\mathbf{k}^{\prime}}$ and $V^{-1}\sum_{\mathbf{k}%
^{\prime\prime}}U_{A}\left(  \mathbf{k},\mathbf{k}^{\prime\prime}\right)
U_{A}^{-1}\left(  \mathbf{k}^{\prime\prime},\mathbf{k}^{\prime}\right)
=\delta_{\mathbf{k},\mathbf{k}^{\prime}}$.

Note that in problems where Bose and Fermi number densities are fixed, the
second term in Eq. (\ref{mu}) (everything inside the square brackets)
represents a constant shift and may thus be absorbed into the chemical
potential $u_{F}$, making it unnecessary to introduce $\mu_{F}$ as we have
done in Eq. (\ref{mu}). \ This, however, no longer holds to be true in the
case of phase separation. \ Phase separation redistributes the particle number
densities among coexisting phases so that the second term in Eq. (\ref{mu}),
and consequently $\mu_{F}$, will be different for different phases, even
though the coexisting phases all share the same chemical potential $u_{F}$.
\ Thus, in order for our theory to handle the subject of phase separation in a
unified manner, we keep the second term explicitly.

To obtain the partition function $Z$ and the thermodynamical potential
$\Omega$, we make use of the Bogoliubov transformation between $\hat
{a}_{\mathbf{k}}$ and the fermionic quansiparticle operator $\hat
{b}_{\mathbf{k}}$: $\hat{a}_{\mathbf{k}}=u_{\mathbf{k}}\hat{b}_{\mathbf{k}%
}+v_{\mathbf{k}}\hat{b}_{-\mathbf{k}}^{\dag}$ and $\hat{a}_{-\mathbf{k}}%
^{\dag}=u_{\mathbf{k}}^{\ast}\hat{b}_{-\mathbf{k}}^{\dag}-v_{\mathbf{k}}%
^{\ast}\hat{b}_{\mathbf{k}}$, where $\left\vert u_{\mathbf{k}}\right\vert
^{2}+$ $\left\vert v_{\mathbf{k}}\right\vert ^{2}=1$, and change the
Hamiltonian $\hat{H}_{F}^{M}$ into the diagonal form
\begin{equation}
\hat{H}_{F}^{M}=\frac{1}{2}\sum_{\mathbf{k}}\left[  \left(  \xi_{\mathbf{k}%
}-E_{\mathbf{k}}\right)  +E_{\mathbf{k}}\left(  \hat{b}_{\mathbf{k}}^{\dag
}\hat{b}_{\mathbf{k}}+\hat{b}_{-\mathbf{k}}^{\dag}\hat{b}_{-\mathbf{k}%
}\right)  \right]  ,
\end{equation}
where $E_{\mathbf{k}}=\sqrt{\xi_{\mathbf{k}}^{2}+\Delta^{2}\left(
\mathbf{k}\right)  }$ is the quansiparticle spectrum of the BCS superfluid.
\ The corresponding grand-cannonical thermodynamic potential (density)
[$\Omega=-\ln Z/(\beta V)$ where $\beta=1/\left(  k_{B}T\right)  $ and $k_{B}$
the Boltzmann constant] is then given by%
\begin{align}
\Omega &  =\frac{1}{V}\left(  H_{B}^{0}+H_{F}^{0}\right)  +\nonumber\\
&  \frac{1}{2V}\sum_{\mathbf{k}}\left(  \xi_{\mathbf{k}}-E_{\mathbf{k}%
}\right)  +\frac{1}{\beta V}\sum_{\mathbf{k}}\ln f\left(  -E_{\mathbf{k}%
}\right)  . \label{Omega}%
\end{align}

The phase transition from a normal gas to the BCS superfluid is known to be
second order in nature. \ In the parameter regime near the second order phase
transition, the gap parameter is small. \ As a result, we may determine the
properties of the Fermi gas close to the second-order phase transition by
expanding the thermodynamical potential up to the fourth order in the gap
parameter. \ We divide this expansion into two parts%
\begin{equation}
\Omega=\Omega_{0}+\Omega_{1}, \label{laudau expansion}%
\end{equation}
where%
\begin{align}
\Omega_{0}  &  =\frac{1}{2}U_{BB}n_{B}^{2}-u_{B}n_{B}-\frac{1}{2}U\left(
0\right)  n_{F}^{2}\nonumber\\
&  +\frac{1}{V}\sum_{\mathbf{k}}\alpha_{0}\left(  \mathbf{k}\right)  +\frac
{1}{2V}\sum_{\mathbf{k},\mathbf{k}^{\prime}}\Sigma\left(  \mathbf{k}\right)
U^{-1}\left(  \mathbf{k}-\mathbf{k}^{\prime}\right)  \Sigma\left(
\mathbf{k}^{\prime}\right)  , \label{Omega0}%
\end{align}
represents the energy if the fermions were all in the normal state, and%
\begin{align}
\Omega_{1}  &  =-\frac{1}{2V}\sum_{\mathbf{k},\mathbf{k}^{\prime}}%
\Delta\left(  \mathbf{k}\right)  \left[  \alpha_{2}\left(  \mathbf{k}\right)
\delta_{\mathbf{k},\mathbf{k}^{\prime}}+U_{A}^{-1}\left(  \mathbf{k}%
,\mathbf{k}^{\prime}\right)  \right]  \Delta^{\ast}\left(  \mathbf{k}^{\prime
}\right) \nonumber\\
&  +\frac{1}{2V}\sum_{\mathbf{k}}\alpha_{4}\left(  \mathbf{k}\right)
\left\vert \Delta\left(  \mathbf{k}\right)  \right\vert ^{4}, \label{Omega1}%
\end{align}
represents the energy contributed by Cooper pairs, where
\begin{align}
\alpha_{0}\left(  \mathbf{k}\right)   &  =\frac{1}{\beta}\ln\left[  f\left(
-\xi_{\mathbf{k}}\right)  \right]  ,\label{alpha0}\\
\alpha_{2}\left(  \mathbf{k}\right)   &  =\frac{1-2f\left(  \xi_{\mathbf{k}%
}\right)  }{2\xi_{\mathbf{k}}}=\frac{\tanh\frac{\beta\xi_{\mathbf{k}}}{2}%
}{2\xi_{\mathbf{k}}},\label{alpha2}\\
\alpha_{4}\left(  \mathbf{k}\right)   &  =\frac{1}{4\xi_{\mathbf{k}}^{2}%
}\left[  \frac{1-2f\left(  \xi_{\mathbf{k}}\right)  }{2\xi_{\mathbf{k}}}-\beta
f\left(  \xi_{\mathbf{k}}\right)  \left(  1-f\left(  \xi_{\mathbf{k}}\right)
\right)  \right]  , \label{alpha4}%
\end{align}
are the relevant expansion coefficients expressed in terms of the Fermi
distribution function $f\left(  x\right)  =\left(  e^{x\beta}+1\right)  ^{-1}%
$. Equation (\ref{laudau expansion}), (\ref{Omega0}) and (\ref{Omega1}) serve
as the foundation for our studies below.

\section{ Critical Temperature when Fock Potential is included}

\label{Sec:surface deformation}

In this section, we discuss how to compute the critical temperature when the
Fock potential is included. \ \ The process begins with the equations for the
Fermi density%
\begin{equation}
n_{F}=\frac{1}{V}\sum_{\mathbf{k}^{\prime}}\frac{1}{2}\left(  1-\tanh
\frac{\beta\xi_{\mathbf{k}^{\prime}}}{2}\right)  , \label{nF}%
\end{equation}
and Fock potential
\begin{equation}
\Sigma\left(  \mathbf{k}\right)  =-\frac{1}{V}\sum_{\mathbf{k}^{\prime}%
}U\left(  \mathbf{k}-\mathbf{k}^{\prime}\right)  \frac{1}{2}\left[
1-\tanh\frac{\beta\xi_{\mathbf{k}^{\prime}}}{2}\right]  , \label{sigma(k)}%
\end{equation}
which are obtained from $n_{F}=-\partial\Omega/\partial\mu_{F}$ and the saddle
point condition for the Fock potential, $\partial\Omega/\partial\Sigma\left(
\mathbf{k}\right)  =0$. \ The two equations are then coupled to the gap
equation
\begin{align}
\Delta\left(  \mathbf{k}\right)   &  =-\frac{1}{V}\sum_{\mathbf{k}^{\prime}%
}U_{A}\left(  \mathbf{k},\mathbf{k}^{\prime}\right)  K\left(  \mathbf{k}%
^{\prime}\right)  \Delta\left(  \mathbf{k}^{\prime}\right) \nonumber\\
&  +\frac{1}{V}\sum_{\mathbf{k}^{\prime}}U_{A}\left(  \mathbf{k}%
,\mathbf{k}^{\prime}\right)  \alpha_{4}\left(  \mathbf{k}^{\prime}\right)
\left\vert \Delta\left(  \mathbf{k}^{\prime}\right)  \right\vert ^{2}%
\Delta\left(  \mathbf{k}^{\prime}\right)  , \label{gap}%
\end{align}
due to the saddle point condition for the gap parameter, $\partial
\Omega/\partial\Delta\left(  \mathbf{k}\right)  =0$. \ In deriving Eq.
(\ref{gap}), we renormalized the dipolar interaction by applying a procedure
similar to the standard method for renormalizing the contact interaction
\cite{bijlsma00}, but expressed in terms of vertex functions \cite{baranov02},
so that%

\begin{equation}
\alpha_{2}\left(  \mathbf{k}\right)  \rightarrow K\left(  \mathbf{k}\right)
=\frac{\tanh\left(  \beta\xi_{\mathbf{k}}/2\right)  }{2\xi_{\mathbf{k}}}%
-\frac{1}{2\epsilon_{\mathbf{k}}}. \label{Kk}%
\end{equation}
Near the critical temperature, the gap is just about to vanish, and we can
ignore the nonlinear term. \ This leads to the linearly coupled homogeneous
equation%
\begin{equation}
\Delta\left(  \mathbf{k}\right)  =\frac{1}{V}\sum_{\mathbf{k}^{\prime}%
}R\left(  \mathbf{k},\mathbf{k}^{\prime}\right)  \Delta\left(  \mathbf{k}%
^{\prime}\right)  , \label{Delta 1}%
\end{equation}
where $R\left(  \mathbf{k},\mathbf{k}^{\prime}\right)  =-U_{A}\left(
\mathbf{k},\mathbf{k}^{\prime}\right)  K\left(  \mathbf{k}^{\prime}\right)
~$is defined as the matrix element of $R$ in momentum space. \ The critical
temperature then corresponds to the largest\ temperature root of the
characteristic (or secular) equation of the matrix $R$:
\begin{equation}
\text{Det}\left(  R-I\right)  =0, \label{charateristic equation}%
\end{equation}
where $I$ is the unit matrix.

In general, one needs to include not only states on the Fermi surface but also
those off the Fermi surface when solving simultaneously Eqs. (\ref{nF}),
(\ref{sigma(k)}) and (\ref{charateristic equation}), which can make
determining the critical temperature a computationally expensive task. In
order to gain quick insight into the critical temperature, we focus on the low
temperature limit where we can make two significant simplifications. First,
Eqs. (\ref{nF}) and (\ref{sigma(k)})\ decouple from the gap equation, and
second, mainly the states near the Fermi surface contribute to pairings. \ As
such, the theory we formulate shall remain quantitatively accurate in the weak
coupling regime and is expected to capture qualitative features of the
critical temperature under more general circumstances.

\begin{figure}
[ptb]
\begin{center}
\includegraphics[
width=3in
]%
{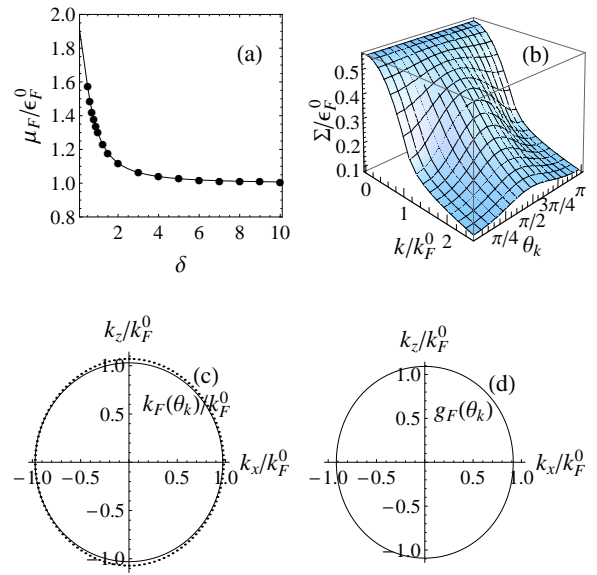}%
\caption{(a) The chemical potential, $\mu_{F}$, is plotted as a function of
the unitless variable $\delta=\xi_{B}k_{F}^{0}$, where $\xi_{B}$ is the
healing length [see Eq. (\ref{delta})]. \ The dotted curve is computed
self-consistently from Eqs. (\ref{nF}) and (\ref{sigma(k)}) and may be
compared to the solid one which is obtained variationally from Eqs.
(\ref{variational E}) and (\ref{uF}). (b) The self-consistently computed Fock
potential, $\Sigma\left(  k,\theta_{\mathbf{k}}\right)  $, for $\delta=0.8$.
(c) A polar plot of the Fermi surface, $k_{F}\left(  \theta_{\mathbf{k}%
}\right)  $. \ The dotted curve is extracted from the Fock potential plotted
in (b) and may be compared to the solid curve which is determined
variationally from Eqs. (\ref{variational E}) and (\ref{K_F variation}). \ (d)
A polar plot of $g_{F}\left(  \theta_{\mathbf{k}}\right)  $, which is defined
in Eq. (\ref{N()}) and is proportional to the density of states at the Fermi
surface. The plot in (d) is determined variationally from Eqs.
(\ref{variational E}) and (\ref{gF(k)}). \ In (c) and (d), the polar angle
$\theta_{\mathbf{k}}$ is measured with respect to the positive $k_{z}$-axis.
In each plot, we used $\lambda=0.87$ and $\varepsilon_{dd}=0.8$, where
$\lambda=N\left(  \epsilon_{F}^{0}\right)  U_{BF}^{2}/U_{BB}$ is defined in
Eq. (\ref{lambda}), and $\varepsilon_{dd}=4\pi d^{2}/(3U_{BB})$ measures the
dipolar interaction relative to the s-wave interaction.}%
\label{Fig1}%
\end{center}
\end{figure}

To proceed, we first set the temperature in Eqs. (\ref{nF}) and
(\ref{sigma(k)}) to zero. In this way, both the chemical and Fock potentials
for a given $n_{F}$ can be obtained self-consistently from Eqs. (\ref{nF}) and
(\ref{sigma(k)}) prior to being fed into the gap equation for estimating the
critical temperature. \ Figure \ref{Fig1} (a) displays the chemical potential
$\mu$ as a function of $\delta$ (dots) when $\varepsilon_{dd}=0.8$ and
$\lambda=0.87\ $and Fig. \ref{Fig1} (b) illustrates how the corresponding Fock
potential $\Sigma\left(  \mathbf{k}\right)  $ changes with $k$ and
$\theta_{\mathbf{k}}$ when $\delta=0.8$, where
\begin{equation}
\delta\equiv\xi_{B}k_{F}^{0}=\frac{k_{F}^{0}}{4\sqrt{\pi n_{B}a_{BB}}},
\label{delta}%
\end{equation}
and
\begin{equation}
\lambda\equiv N\left(  \epsilon_{F}^{0}\right)  U_{BF}^{2}/U_{BB}=\frac{2}%
{\pi}\frac{m_{B}m_{F}}{m_{BF}^{2}}\frac{a_{BF}^{2}}{a_{BB}}k_{F}^{0},
\label{lambda}%
\end{equation}
with
\begin{equation}
N\left(  \epsilon_{F}^{0}\right)  =\frac{m_{F}k_{F}^{0}}{2\pi^{2}\hbar^{2}}
\label{density of states}%
\end{equation}
the density of states at the Fermi surface. The use of the superscript $0$
here is to stress that
\begin{equation}
k_{F}^{0}=\left(  6\pi^{2}n_{F}\right)  ^{1/3},\text{ }\epsilon_{F}^{0}%
=\hbar^{2}\left(  k_{F}^{0}\right)  ^{2}/2m_{F}, \label{Fermi wavenumber}%
\end{equation}
are, respectively, the Fermi energy and wavenumber of a spin-polarized
\textit{non-interacting }Fermi gas with number density $n_{F}$. \ Note that
the rotational symmetry of the induced interaction with respect to the $z$
axis implies that $U\left(  \mathbf{k}-\mathbf{k}^{\prime}\right)  $ is a
function of the relative azimuthal angle $\phi_{\mathbf{k}}-\phi
_{\mathbf{k}^{\prime}}$, and \ physical observables such as $\Sigma\left(
\mathbf{k}\right)  $, being only dependent on the induced interaction after
the azimuthal degree of freedom is integrated out in the manner
\begin{equation}
\int_{0}^{2\pi}U\left(  \mathbf{k}-\mathbf{k}^{\prime}\right)  d\left(
\phi_{\mathbf{k}}-\phi_{\mathbf{k}^{\prime}}\right)  , \label{average U}%
\end{equation}
are thus functions of $\left(  k,\theta_{\mathbf{k}}\right)  $ only. \ As a
result, $\phi_{\mathbf{k}}$ in $\Sigma\left(  \mathbf{k}\right)  $ [of
Fig.\ \ref{Fig1}(b)] has been suppressed; this convention is to be extended to
all the other observables. Additionally, for notational simplicity, $U\left(
\mathbf{k}-\mathbf{k}^{\prime}\right)  $ and $U_{A}\left(  \mathbf{k}%
,\mathbf{k}^{\prime}\right)  $ are to be understood, from now on, to represent
the ones where azimuthal coordinates have been integrated out as in Eq.
(\ref{average U}). \ Figure 1(b) indicates that the Fock potential is lower in
the external field direction ($\theta_{\mathbf{k}}=0$ and $\pi$ or the z-axis)
than in the orthogonal direction ($\theta_{\mathbf{k}}=\pi/2$) leading to a
Fermi surface slightly elongated along the dipole orientation as shown in Fig.
1(c) (dotted curve), a feature apparently stemming from the anisotropic nature
of the induced interaction in Eq. (\ref{Uind}).%

%EndExpansion

\bigskip To capture the anisotropic nature of the Fock potential, we organize
the single-particle dispersion
\begin{equation}
\xi_{\mathbf{k}}=\frac{\hbar^{2}}{2m_{F}\gamma\left(  k,\theta_{\mathbf{k}%
}\right)  }\left[  k^{2}-k_{F}^{2}\left(  \theta_{\mathbf{k}}\right)  \right]
, \label{dispersion}%
\end{equation}
in terms of two parameters, $k_{F}\left(  \theta_{\mathbf{k}}\right)  $ and
$\gamma\left(  k,\theta_{\mathbf{k}}\right)  $, which, in principle, can
always be extracted from the numerically computed Fock potential.
\ \ Physically, the former defines the Fermi surface $k=k_{F}\left(
\theta_{\mathbf{k}}\right)  $, while the latter allows us to define
$m_{F}\gamma_{F}\left(  \theta_{\mathbf{k}}\right)  $ as the renormalized mass
along the $\theta_{\mathbf{k}}$ direction at the Fermi surface, where
$\gamma_{F}\left(  \theta_{\mathbf{k}}\right)  =\lim_{k\rightarrow
k_{F}\left(  \theta_{\mathbf{k}}\right)  }\gamma\left(  k,\theta_{\mathbf{k}%
}\right)  $. \ These same physical interpretations also follow from the roles
these quantities play in modifying the density of states. \ By definition, the
number of states per unit volume, $dN$, along $\theta_{\mathbf{k}}$ within the
solid differential angle $d\Omega_{\mathbf{k}}=\sin\theta_{\mathbf{k}}%
d\theta_{\mathbf{k}}d\phi_{\mathbf{k}}$, is given by
\begin{equation}
dN=\frac{1}{(2\pi)^{3}}\frac{k^{2}}{\left\vert \nabla_{\mathbf{k}}%
\xi_{\mathbf{k}}\right\vert }d\Omega_{\mathbf{k}}\equiv N\left(
\Omega_{\mathbf{k}}\right)  \frac{d\Omega_{\mathbf{k}}}{4\pi}, \label{dN}%
\end{equation}
where $N\left(  \Omega_{\mathbf{k}}\right)  $ is defined as the density of
states. \ This\ means that in the limit where the Fermi surface varies slowly
with $\theta_{\mathbf{k}}$ in the sense that $\left\vert k_{F}^{-1}\left(
\theta_{\mathbf{k}}\right)  dk_{F}\left(  \theta_{\mathbf{k}}\right)
/d\theta_{\mathbf{k}}\right\vert \ll1$, the density of states at the Fermi
surface,
\begin{equation}
N\left(  \Omega_{\mathbf{k}}\right)  =N\left(  \epsilon_{F}^{0}\right)
\left[  g_{F}\left(  \theta_{\mathbf{k}}\right)  \equiv\gamma_{F}\left(
\theta_{\mathbf{k}}\right)  k_{F}\left(  \theta_{\mathbf{k}}\right)
/k_{F}^{0}\right]  , \label{N()}%
\end{equation}
simply corresponds to Eq. (\ref{density of states})\ for the isotropic case in
which $k_{F}^{0}$ and $m_{F}$ are replaced, respectively, with $k_{F}\left(
\theta_{\mathbf{k}}\right)  $ and $m_{F}\gamma_{F}\left(  \theta_{\mathbf{k}%
}\right)  $. \ This result is in complete agreement with our expectation in
light of the discussion above. \ Thus, we see that the Fock potential
introduces anisotropy in the density of states so that the density of states
at the Fermi surface is modulated along the angular dimension by a factor of
$g_{F}\left(  \theta_{\mathbf{k}}\right)  $.

In the low temperature limit $k_{B}T/\epsilon_{F}^{0}\ll1$, $k^{2}K\left(
\mathbf{k}\right)  $ is negligible except around the Fermi surface
$k=k_{F}\left(  \theta_{\mathbf{k}}\right)  $, where it is sharply peaked
compared to other momentum distributions inside the integration kernel in Eq.
(\ref{Delta 1}). \ This allows us to apply an analogous procedure, well-known
in the study of isotropic BCS pairing \cite{landau80}, to single out the
states near the Fermi surface as the dominant contribution to the gap
equation. \ Indeed, we find, to a good approximation (correct up to a
pre-exponential factor), that the gap at the Fermi surface $\Delta\left(
\theta_{\mathbf{k}}\right)  \equiv\Delta\left(  k,\theta_{\mathbf{k}}\right)
|_{k=k_{F}\left(  \theta_{\mathbf{k}}\right)  }$ obeys the equation
\begin{align}
\Delta\left(  \theta_{\mathbf{k}}\right)   &  \mathbf{=}\mathbf{-}%
\frac{\lambda}{4\pi}\ln\frac{\pi k_{B}T}{8\epsilon_{F}^{0}e^{\gamma-2}}%
\times\nonumber\\
&  \int g_{F}\left(  \theta_{\mathbf{k}^{\prime}}\right)  \bar{U}_{A}\left(
\theta_{\mathbf{k}},\theta_{\mathbf{k}^{\prime}}\right)  \Delta\left(
\theta_{\mathbf{k}^{\prime}}\right)  \sin\theta_{\mathbf{k}^{\prime}}%
d\theta_{\mathbf{k}^{\prime}}, \label{Delta Theda}%
\end{align}
where
\begin{equation}
\bar{U}_{A}\left(  \theta_{\mathbf{k}},\theta_{\mathbf{k}^{\prime}}\right)
\equiv\frac{U_{A}\left(  k,\theta_{\mathbf{k}},k^{\prime},\theta
_{\mathbf{k}^{\prime}}\right)  |_{k=k_{F}\left(  \theta_{\mathbf{k}}\right)
,k^{\prime}=k_{F}\left(  \theta_{\mathbf{k}^{\prime}}\right)  }}{U\left(
0\right)  },
\end{equation}
is the matrix element of the scaled interaction between two momenta on the
Fermi surface. \ The critical temperature then becomes%
\begin{equation}
T=\frac{8\epsilon_{F}^{0}e^{\gamma-2}}{\pi k_{B}}\exp\left[  -\frac{1}%
{\omega\lambda}\right]  , \label{Tn}%
\end{equation}
where $\omega$ is the largest positive eigenvalue of the eigenvalue equation
\begin{equation}
4\pi\int g_{F}\left(  \theta_{\mathbf{k}^{\prime}}\right)  \bar{U}_{A}\left(
\theta_{\mathbf{k}},\theta_{\mathbf{k}^{\prime}}\right)  \psi\left(
\theta_{\mathbf{k}^{\prime}}\right)  \sin\theta_{\mathbf{k}^{\prime}}%
d\theta_{\mathbf{k}^{\prime}}=\omega\psi\left(  \theta_{\mathbf{k}}\right)  .
\label{charateristic equation 1}%
\end{equation}
As can be seen, when we set $g_{F}\left(  \theta_{\mathbf{k}^{\prime}}\right)
=1$, we recover from Eqs. (\ref{Delta Theda}) and
(\ref{charateristic equation 1}) the corresponding equations in the absence of
Fermi surface deformation.

Inspired by the work in \cite{miyakawa08,fregoso09} and the fact that our
numerically determined Fermi surface fits well with an ellipsoid, we assume
that the Femi particle number $n_{\mathbf{k}}\equiv\left\langle \hat
{a}_{\mathbf{k}}^{\dag}\hat{a}_{\mathbf{k}}\right\rangle $ distributes
according to the variational ansatz
\begin{equation}
n_{\mathbf{k}}=\Theta\left[  \left(  k_{F}^{0}\right)  ^{2}-\alpha^{-1}\left(
k_{x}^{2}+k_{y}^{2}\right)  -\alpha^{2}k_{z}^{2}\right]  ,
\label{number distribution}%
\end{equation}
where $\Theta\left[  x\right]  $ is the Heaviside step function, and $\alpha$
is the so-called deformation parameter; the Fermi surface is prolate when
$\alpha<1$ and oblate when $\alpha>1$. \ An explicit evaluation of both
kinetic energy and Fock potential indicates that $\alpha$ may be determined
variationally by minimizing the (kinetic + Fock exchange) energy density:
\begin{equation}
\mathcal{E}\left(  \alpha\right)  =\frac{1}{2}\frac{\hbar^{2}}{m_{F}}%
\frac{\left(  k_{F}^{0}\right)  ^{5}}{10\pi^{2}}Y\left(  \alpha\right)  ,
\label{variational E}%
\end{equation}
where
\begin{align}
Y\left(  \alpha\right)   &  =\frac{2}{3}\alpha+\frac{1}{3\alpha^{2}%
}\nonumber\\
&  +\frac{5}{12}\lambda\text{ }\int_{0}^{\pi}I\left(  \theta_{\mathbf{k}%
},\alpha,\varepsilon_{dd},\delta\right)  \sin\theta_{\mathbf{k}}%
d\theta_{\mathbf{k}}. \label{Y}%
\end{align}
In Eq. (\ref{Y}),
\begin{align}
&  I\left(  \theta_{\mathbf{k}},\alpha,\varepsilon_{dd},\delta\right)
\nonumber\\
&  =\frac{1}{4d}\left(  6-\frac{c}{d}\right)  -\frac{2}{d}\sqrt{\frac{c}{d}%
}\tan^{-1}\left(  2\sqrt{\frac{d}{c}}\right) \nonumber\\
&  +\frac{1}{16d}\frac{c}{d}\left(  \frac{c}{d}+12\right)  \ln\left(
1+4\frac{d}{c}\right)
\end{align}
plays the same role as the deformation function in dipolar Fermi gases
\cite{miyakawa08}, but with a more complicated form that depends not only on
the deformation parameter $\alpha$ but also on $\varepsilon_{dd}$ and $\delta
$, \ where
\begin{align}
c  &  =1-\varepsilon_{dd}+3\varepsilon_{dd}\frac{\cos^{2}\theta_{\mathbf{k}}%
}{\alpha^{3}\sin^{2}\theta_{\mathbf{k}}+\cos^{2}\theta_{\mathbf{k}}}\text{
,}\\
d  &  =\frac{\delta^{2}}{\alpha^{2}}\left(  \alpha^{3}\sin^{2}\theta
_{\mathbf{k}}+\cos^{2}\theta_{\mathbf{k}}\right)  .
\end{align}
For an ellipsoid whose Fermi surface is defined by Eq.
(\ref{number distribution}), \
\begin{align}
k_{F}\left(  \theta_{\mathbf{k}}\right)   &  \equiv k_{F}^{0}\left(
\alpha^{-1}\sin^{2}\theta_{\mathbf{k}}+\alpha^{2}\cos^{2}\theta_{\mathbf{k}%
}\right)  ^{-1/2},\label{K_F variation}\\
\gamma_{F}\left(  \theta_{\mathbf{k}}\right)   &  =\left(  \alpha^{-1}\sin
^{2}\theta_{\mathbf{k}}+\alpha^{2}\cos^{2}\theta_{\mathbf{k}}\right)  ^{-1},
\end{align}
and consequently according to Eq. (\ref{N()})%
\begin{equation}
g_{F}\left(  \theta_{\mathbf{k}}\right)  =\left(  \alpha^{-1}\sin^{2}%
\theta_{\mathbf{k}}+\alpha^{2}\cos^{2}\theta_{\mathbf{k}}\right)  ^{-3/2}.
\label{gF(k)}%
\end{equation}
The corresponding chemical potential, calculated using $\mu_{F}=\partial
\mathcal{E}/\partial n_{F}$, is found to be
\begin{align}
\frac{\mu_{F}}{\epsilon_{F}^{0}}  &  =\frac{6}{5}Y\left(  \alpha\right)
-\frac{1}{5}\left(  \frac{2}{3}\alpha+\frac{1}{3\alpha^{2}}\right) \nonumber\\
&  +\frac{1}{12}\lambda\int_{0}^{\pi}K\left(  \theta_{\mathbf{k}}%
,\alpha,\varepsilon_{dd},\delta\right)  \sin\theta_{\mathbf{k}}d\theta
_{\mathbf{k}}, \label{uF}%
\end{align}
where%
\begin{align}
&  K\left(  \theta_{\mathbf{k}},\alpha,\varepsilon_{dd},\delta\right)
\nonumber\\
&  =3\left(  \frac{c}{2d^{2}}-\frac{1}{d}\right)  +6\frac{\sqrt{c}}{d^{3/2}%
}\tan^{-1}\left(  2\frac{d^{1/2}}{\sqrt{c}}\right) \nonumber\\
&  -\frac{c}{8}\left(  \frac{3c}{d^{3}}+\frac{24}{d^{2}}\right)  \ln\left(
1+\frac{4}{c}d\right)  .
\end{align}

Figure 1(a) shows that Eq. (\ref{uF}) produces a chemical potential (solid
curve) that is in good agreement with the one obtained self-consistently from
Eqs. (\ref{nF}) and (\ref{sigma(k)}) (dots). \ Figure 1(c) shows that compared
to the self-consistent method (dotted curve), the variational method (black
curve) underestimates the prolateness of the Fermi surface, but nonetheless
captures the essential anisotropic feature of the Fermi surface and does not,
in our judgement, prevent us from using it as a computationally economic tool
for estimating the critical temperature.

\begin{figure}
[ptb]
\begin{center}
\includegraphics[
width=3in
]%
{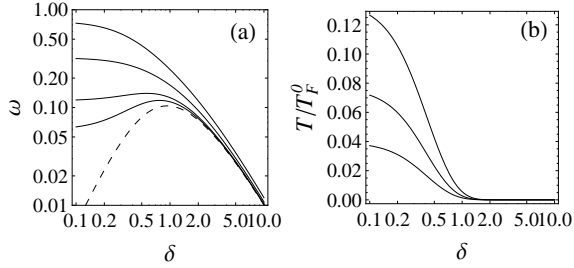}%
\caption{The superfluid critical temperature, $T$, is proportional to
$\exp\left(  -1/\omega\lambda\right)  $ [Eq. (\ref{Tn})], where $\omega$ is
the largest positive eigenvalue of Eq. (\ref{charateristic equation 1}) and
$\lambda=N\left(  \epsilon_{F}^{0}\right)  U_{BF}^{2}/U_{BB}$ is defined in
Eq. (\ref{lambda}). \ (a) The eigenvalue, $\omega$, is plotted as a function
of the unitless variable $\delta=\xi_{B}k_{F}^{0}$, where $\xi_{B}$ is the
healing length [see Eq. (\ref{delta})]. \ The dashed curve corresponds to the
absence of the dipolar interaction ($\varepsilon_{dd}=0$). \ The solid curves,
from bottom to top, correspond to $\varepsilon_{dd}$ = $0.1,$ $0.2$, 0.5 and
0.8, where $\varepsilon_{dd}=4\pi d^{2}/(3U_{BB})$ measures the dipolar
interaction relative to the s-wave interaction. (b) The critical temperature,
$T$, in the units of $T_{F}^{0}\equiv\epsilon_{F}^{0}/k_{B}$ is plotted as a
function of $\delta$. \ The curves, from bottom to top, correspond to
$\varepsilon_{dd}$=0.6, 0.7 and 0.8. \ All curves are produced with
$\lambda=0.87$.}%
\label{Fig2}%
\end{center}
\end{figure}

Our algorithm for computing the critical temperature is then as follows: we
first calculate $\alpha$ by minimizing Eq. (\ref{variational E}) with respect
to $\alpha$; next we use it to fix $g_{F}\left(  \theta_{\mathbf{k}}\right)  $
according to Eq. (\ref{gF(k)}); we then substitute $g_{F}\left(
\theta_{\mathbf{k}}\right)  $ into Eq. (\ref{charateristic equation 1}) to
determine the eigenvalue $\omega$; finally we use Eq. (\ref{Tn}) to estimate
the critical temperature. \ Figure \ref{Fig2} (a) compares the eigenvalue in
the absence of dipolar interaction (dashed curve) with those in the presence
of dipolar interaction (solid curves), indicating that in contrast to the
former, which asymptotes to zero, in the limit of small $\delta$, the latter
approach finite values, which increase dramatically with the increase in
$\varepsilon_{dd}$. \ This is explicitly shown in Fig. \ref{Fig2} (b), which
is a plot of the critical temperature for various $\varepsilon_{dd}$. \ Not
shown in Fig. \ref{Fig2}(b) is the critical temperature in the absence of the
dipolar interaction, which is orders of magnitude lower.%
%TCIMACRO{\FRAME{ftbpFU}{3.2292in}{1.7201in}{0pt}{\Qcb{The superfluid critical
%temperature, $T$, is proportional to $\exp\left(  -1/\omega\lambda\right)  $
%[Eq. (\ref{Tn})], where $\omega$ is the largest positive eigenvalue of Eq.
%(\ref{charateristic equation 1}) and $\lambda=N\left(  \epsilon_{F}%
%^{0}\right)  U_{BF}^{2}/U_{BB}$ is defined in Eq. (\ref{lambda}). \ (a) The
%eigenvalue, $\omega$, is plotted as a function of the unitless variable
%$\delta=\xi_{B}k_{F}^{0}$, where $\xi_{B}$ is the healing length [see Eq.
%(\ref{delta})]. \ The dashed curve corresponds to the absence of the dipolar
%interaction ($\varepsilon_{dd}=0$). \ The solid curves, from bottom to top,
%corresponds to $\varepsilon_{dd}$ = $0.1,$ $0.2$, 0.5 and 0.8, where
%$\varepsilon_{dd}=4\pi d^{2}/(3U_{BB})$ measures the dipolar interaction
%relative to the s-wave interaction. (b) The critical temperature, $T$, in the
%unit of $T_{F}^{0}\equiv\epsilon_{F}^{0}/k_{B}$ is plotted as a function of
%$\delta$. \ The curves correspond to, from bottom to top, $\varepsilon_{dd}%
%$=0.6, 0.7 and 0.8. \ All curves are produced with $\lambda=0.87$.}%
%}{\Qlb{Fig2}}{fig2.eps}{\special{ language "Scientific Word";
%type "GRAPHIC";  maintain-aspect-ratio TRUE;  display "USEDEF";
%valid_file "F";  width 3.2292in;  height 1.7201in;  depth 0pt;
%original-width 6.0027in;  original-height 3.186in;  cropleft "0";
%filename 'fig2.eps';file-properties "XNPEU";}}}%
%BeginExpansion
%EndExpansion

\section{Optimal critical temperature prior to phase separation}

\label{Sec:optimization}

The issue of whether different phases can coexist in the same spatial volume
(miscibility) or repel each other into separate spatial regions (phase
separation) will inevitably arise when one is considering systems made up of
quantum gases with different species. \ The knowledge of phase separation is
essential before one can apply the theory developed in the previous sections
to estimate the critical temperature achievable before the system starts to
phase separate. An understanding of the phase separation, even in\ view of the
tremendous simplification due to the use of the variational method, would most
likely\ remain a computationally expensive endeavor if one were to take into
full consideration the Fock potential. \ Consequently, when studying phase
separation we choose to ignore the momentum dependence of the induced
interaction by assuming $U\left(  \mathbf{k}\right)  =U\left(  0\right)  $.
\ The modification due to this assumption is not expected to undermine our
goal (in this section) of gaining a qualitative understanding about the order
of magnitude of the achievable critical temperature. \ This assumption allows
the Fock potential to cancel the Hartree potential identically, leading to a
much simplified $\Omega$, which, when expressed in terms of $A=$ $\hbar
^{2}\left(  6\pi^{2}\right)  ^{2/3}/2m_{F}$, takes the form
\begin{align}
\Omega &  =\frac{1}{2}U_{BB}n_{B}^{2}-u_{B}n_{B}+U_{BF}n_{B}n_{F}\nonumber\\
&  +\frac{3}{5}An_{F}^{5/3}-u_{F}n_{F}, \label{Omega0 1}%
\end{align}
where we have used Eqs. (\ref{Fock}), (\ref{Omega0}), (\ref{alpha0}) and
(\ref{nF}), and ignored $\Omega_{1}$, the energy contribution from Cooper
pairs, as we have done in other sections.

Note that Eq. (\ref{Omega0 1}) has the same mathematical form as the one for a
homogeneous Fermi-Bose mixture model, which has been extensively analyzed in
\cite{efremov02,viverit00}. \ Here, we recapture the main physics by
performing the same analysis but in the chemical potential space
\cite{fodor10}. Consider first the mixed phase where both $n_{B}$ and $n_{F}$
have finite values. The corresponding chemical potentials are given by \
\begin{align}
u_{B}  &  =\frac{\partial\Omega}{\partial n_{B}}=U_{BB}n_{B}+U_{BF}n_{F},\\
u_{F}  &  =\frac{\partial\Omega}{\partial n_{F}}=U_{BF}n_{B}+An_{F}^{2/3}.
\end{align}
Eliminating the boson density, we arrive at a cubic equation for $n_{F}^{1/3}%
$,%
\begin{equation}
\frac{U_{BF}^{2}}{U_{BB}}\left(  n_{F}^{1/3}\right)  ^{3}-A\left(  n_{F}%
^{1/3}\right)  ^{2}+u_{F}-\frac{U_{BF}}{U_{BB}}u_{B}=0, \label{cubic}%
\end{equation}
which allows us to determine $n_{F}^{1/3}$ given a set of $u_{F}$ and $u_{B}$.
Not all positive solutions to Eq. (\ref{cubic}) are stable and therefore
physically realizable; the values of $n_{F}^{1/3}$ are constrained by the
stability condition which requires the following Hessian matrix to be positive
semidefinite:
\begin{equation}
\left(
\begin{array}
[c]{cc}%
\frac{\partial^{2}\Omega}{\partial n_{B}^{2}} & \frac{\partial^{2}\Omega
}{\partial n_{B}\partial n_{F}}\\
\frac{\partial^{2}\Omega}{\partial n_{F}\partial n_{B}} & \frac{\partial
^{2}\Omega}{\partial n_{F}^{2}}%
\end{array}
\right)  =\left(
\begin{array}
[c]{cc}%
U_{BB} & U_{BF}\\
U_{BF} & A\frac{2}{3}\frac{1}{n_{F}^{1/3}}%
\end{array}
\right)  ,
\end{equation}
or equivalently
\begin{equation}
n_{F}^{1/3}<A\frac{2U_{BB}}{3U_{BF}^{2}}. \label{stability condition}%
\end{equation}
There exists only one positive root to the cubic equation (\ref{cubic}) that
satisfies the stability condition and this happens in the chemical potential
space when
\begin{equation}
\frac{U_{BF}}{U_{BB}}u_{B}<u_{F}<\frac{U_{BF}}{U_{BB}}u_{B}+\frac{4}{27}%
\frac{A^{3}U_{BB}^{2}}{U_{BF}^{4}}.
\end{equation}
This means that there is a unique mixed state, which precludes the possibility
of phase separation involving more than one mixed phase. \ 

Next, \ in the region that supports a pure Bose phase, the effective chemical
potential for fermions shall be less than zero, $u_{F}-U_{BF}n_{B}<0$, which,
when combined with $u_{B}=U_{BB}n_{B}$ for a pure Bose state, shows
immediately that the pure Bose phase exists in the chemical potential space
with
\begin{equation}
u_{F}<U_{BF}u_{B}/U_{BB}.
\end{equation}
Thus, we see that there is no overlap between the pure Bose and the mixed
phase; the two phases are divided by the line $u_{F}=U_{BF}u_{B}/U_{BB}$,
which is second order in nature. \ As a result, phase separation between the
mixed state and a pure Bose state is impossible.

The only remaining possibility is phase separation between a mixed state with
densities $\left(  n_{F1},n_{B1}\right)  $ and a pure Fermi phase with
densities $\left(  n_{F2},n_{B2}=0\right)  $. \ (A complete separation between
fermions and bosons requires much higher densities than considered in the
present work.) \ The mixed phase must share the same chemical and
thermodynamical potentials with the pure phase, meaning%
\begin{align}
U_{BF}n_{B1}+An_{F1}^{2/3}  &  =An_{F2}^{2/3},\label{chem}\\
U_{BB}n_{B1}^{2}/2+U_{BF}n_{B1}n_{F1}+\frac{2}{5}An_{F1}^{5/3}  &  =\frac
{2}{5}An_{F2}^{5/3}, \label{thermo}%
\end{align}
from which we find
\begin{equation}
n_{B1}=\frac{A}{U_{BF}}n_{F1}^{2/3}\left(  y^{2}-1\right)  , \label{nB1}%
\end{equation}
where $y=\left(  n_{F2}/n_{F1}\right)  ^{1/3}$ is the solution to the
equation
\begin{equation}
-\frac{15}{\lambda}\left(  y+1\right)  ^{2}+8y^{3}+16y^{2}+24y+12=0\text{,}
\label{y}%
\end{equation}
where $\lambda$ is given by Eq. (\ref{lambda}) except that $\epsilon_{F}^{0}$
is replaced with the Fermi energy $\epsilon_{F1}^{0}$ of the mixed phase.
\ The stability condition in Eq. (\ref{stability condition}) then means that
the stable mixture takes place when $\lambda$ $<1$.

\begin{figure}
[ptb]
\begin{center}
\includegraphics[
width=3in
]%
{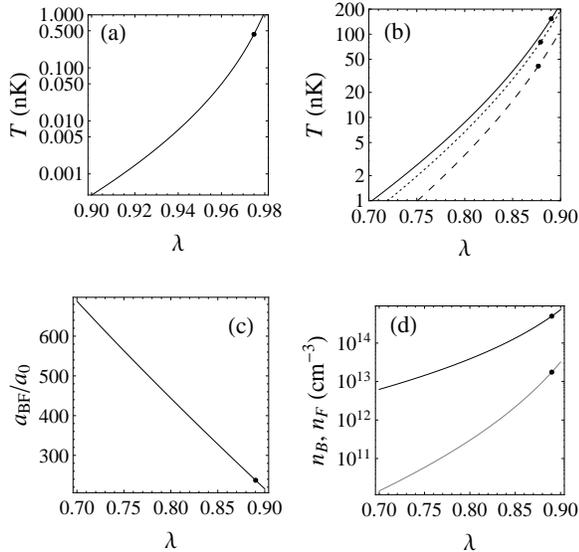}%
\caption{(a) In the absence of the dipolar interaction ($\varepsilon_{dd}=0$),
the superfluid critical temperature, $T$, is plotted as a function of the
unitless variable $\lambda$, where $\lambda=N\left(  \epsilon_{F}^{0}\right)
U_{BF}^{2}/U_{BB}$ is defined in Eq. (\ref{lambda}). (b) For a nonzero dipolar
interaction the critical temperature, $T$, is plotted for $\varepsilon_{dd}=0.6$
(long dash), 0.7 (short dash), and 0.8 (solid), where $\varepsilon_{dd}=4\pi
d^{2}/(3U_{BB})$ measures the dipolar interaction relative to the s-wave
interaction. \ (c) $a_{BF}$ and (d) densities $n_{B}$ [black curve in (d)] and
$n_{F}$ [gray curve in (d)] that are required to produce the (optimal)
critical temperature for $\varepsilon_{dd}=0.8$, represented by the solid
curve in (b). \ The black dots mark the points at which $n_{B}$ has reached
$5\times10^{14}$ cm$^{-3}$. \ All curves are produced with $m_{F}=6$ amu,
$m_{B}=127$ amu, and $a_{BB}=250a_{0}$.}%
\label{Fig3}%
\end{center}
\end{figure}

Let us now turn to the question of how to identify the parameter space which
optimizes the critical temperature before the system phase separates. \ A
possible solution can be found in Ref. \cite{efremov02}. \ Here, we offer an
answer from a different perspective. To be concrete, we consider a system in
which $m_{F}=6$ amu, $m_{B}=127$ amu, $a_{BB}=250a_{0}$ (with amu being the
atomic mass unit and $a_{0}$ the Bohr radius). Our task here is to find, for a
given $\lambda$ and $\varepsilon_{dd}$, a set of $\left(  n_{F,}n_{B}%
,a_{BF}\right)  $ which optimize the critical temperature before the system
phase separates. \ We begin with Eq. (\ref{Tn}), the equation for the critical
temperature. \ We replace $\epsilon_{F}^{0}$ in Eq. (\ref{Tn}) in favor of
$\delta$ defined in Eq. (\ref{delta}) and reorganize Eq. (\ref{Tn}) in terms
of $\delta$ as
\begin{equation}
T=e^{\gamma-2}\frac{8^{2}\hbar^{2}n_{B}a_{BB}}{k_{B}m_{F}}\left\{  \tilde
{T}\left(  \delta\right)  =\delta^{2}\exp\left[  -\frac{1}{\omega\left(
\delta\right)  \lambda}\right]  \right\}  , \label{TT}%
\end{equation}
where the argument in\ $\omega\left(  \delta\right)  $ is to stress that when
$\lambda$ and $\varepsilon_{dd}$ \ are fixed, the eigenvalue $\omega$ is a
function of $\delta$ only. \ On the one hand, in the limit of large $\delta$,
$\omega\left(  \delta\right)  $ is a small (positive) number [See Fig.
\ref{Fig2}(a)], and thus $\tilde{T}\left(  \delta\right)  $ will be low due to
the exponential factor in Eq. (\ref{TT}). On the other hand, in the limit of
small $\delta$, $\tilde{T}\left(  \delta\right)  $ is also low due to the
factor $\delta^{2}$ in Eq. (\ref{TT}). \ \ Thus, \ there will exist some
$\delta=$ $\delta_{peak}$ at which $\tilde{T}\left(  \delta\right)  $ reaches
a maximum value, $\tilde{T}\left(  \delta_{peak}\right)  $. \ Once $\delta$ is
fixed to $\delta_{peak}$, the temperature $T$ according to Eq. (\ref{TT})
increases with the boson density $n_{B}$, \ and is highest when the largest
$n_{B}$\ is used. \ Provided that $n_{B}$ does not exceed values where
three-body recombination begins to dominate the loss mechanism, the
highest\ possible $n_{B}$ before phase separation\ equals $n_{B1}$ in Eq.
(\ref{nB1}) when $n_{F1}=n_{F}$. \ Finally, by combining Eq. (\ref{nB1}) with
Eqs. (\ref{delta}), (\ref{lambda}) and (\ref{Fermi wavenumber}), and solving
them simultaneously, we find that the required $\left(  n_{F,}n_{B}%
,a_{BF}\right)  $ are given by
\begin{align}
n_{F}  &  =\frac{\pi}{6\left(  8^{3}\right)  }\frac{\lambda^{3}\left(
m_{F}/m_{B}\right)  ^{3}}{a_{BB}^{3}\delta_{peak}^{12}\left(  y^{2}-1\right)
^{6}},\\
n_{B}  &  =\frac{\pi}{2\left(  8^{3}\right)  }\frac{\lambda^{2}\left(
m_{F}/m_{B}\right)  ^{2}}{a_{BB}^{3}\delta_{peak}^{10}\left(  y^{2}-1\right)
^{4}},\\
a_{BF}  &  =2a_{BB}\delta_{peak}^{2}\frac{m_{BF}}{m_{F}}\left(  y^{2}%
-1\right)  ,
\end{align}
where $y$ is the solution to Eq. (\ref{y}) for the given $\lambda$. The
results of this optimization procedure are summarized in Fig. \ref{Fig3}. \ A
comparison between Figs. \ref{Fig3}(a) and (b) indicates that mixing fermions
with a dipolar condensate can result in a dramatic increase in the achievable
temperature $T$ compared to mixing fermions with a nondipolar condensate; the
order of magnitude in T can be 100 nK for the former and only a fraction of 1
nK for the latter. \ Figures \ref{Fig3}(c) and (d) display, respectively, the
$a_{BF}$ and $\left(  n_{B},n_{F}\right)  $ required to realize the
temperature in Fig. \ref{Fig3}(b) when $\varepsilon_{dd}=0.8$ (solid curve).
\ For example, when $\lambda=0.874$, a system with $a_{BF}=275a_{0}$,
$n_{B}=2.76\times10^{14}$ cm$^{-3}$ and $n_{F}=6.88\times10^{12}$ cm$^{-3}$
has a critical temperature of $80$ nK. \ \ \
%TCIMACRO{\FRAME{ftbpFU}{2.9897in}{2.9897in}{0pt}{\Qcb{(a) In the absence of
%the dipolar interaction ($\varepsilon_{dd}=0$), the superfluid critical
%temperature, $T$, is plotted as a function of the unitless variable $\lambda$,
%where $\lambda=N\left(  \epsilon_{F}^{0}\right)  U_{BF}^{2}/U_{BB}$ is defined
%in Eq. (\ref{lambda}). (b) For a nonzero dipolar interaction the critical
%temperature, T, is plotted for $\varepsilon_{dd}=0.6$ (long dash), 0.7 (short
%dash), and 0.8 (solid), where $\varepsilon_{dd}=4\pi d^{2}/(3U_{BB})$ measures
%the dipolar interaction relative to the s-wave interaction. \ (c) $a_{BF}$ and
%(d) densities $n_{B}$ [black curve in (d)] and $n_{F}$ [gray curve in (d)]
%that are required to produce the (optimal) critical temperature for
%$\varepsilon_{dd}=0.8$, represented by the solid curve in (b). \ The black
%dots mark the points at which $n_{B}$ has reached $5\times10^{14}$ cm$^{-3}$.
%\ All curves are produced with $m_{F}=6$ amu, $m_{B}=127$ amu, and
%$a_{BB}=250a_{0}$.}}{\Qlb{Fig3}}{fig3.eps}%
%width 2.9897in;  height 2.9897in;  depth 0pt;  original-width 6.0027in;
%original-height 6.0027in;  cropleft "0";  croptop "1";  cropright "1";
%cropbottom "0";  filename 'fig3.eps';file-properties "XNPEU";}}}%
%BeginExpansion
%EndExpansion

\section{Conclusion}

\label{Sec:conclusion}

The ability to easily mix cold atoms of different species opens up another
exciting avenue for engineering quantum systems with novel properties. Mixing
a dipolar condensate can induce, between two fermions in a Fermi quantum gas,
an effective attractive interaction which, owning to its anisotropic nature,
can dramatically enhance the scattering of fermions in the odd-parity
channels. \ We have explored such an interaction for the purpose of achieving,
in a mixture between a spin-polarized Fermi gas and a dipolar condensate, a
superfluid with a gap parameter characterized with a coherent superposition of
all odd partial waves. \ We have focused our effort on determining the
critical temperature, a task far more challenging for our model than for pure
fermionic systems since, first, adding dipolar bosons to fermions dramatically
increases the system parameter space and, second, the system may phase
separate. \ We have formulated, in the spirit of the Hartree-Fock-Bogoliubov
mean-field approach, a theory which allows us to treat calculating the
critical temperature and phase separation in a unified manner. \ \ We have
applied this theory to estimate the critical temperature when the anisotropic
Fock potential is taken into consideration and to identify the parameter space
which optimizes the critical temperature before the system begins to phase separate.

\section{Acknowledgement\ }

H. Y. L. acknowledges the support from the US National Science Foundation and
the US Army Research Office.

\bigskip%

\end{document}